\documentclass[12pt]{spieman}  % 12pt font required by SPIE;
\usepackage{amsmath,amsfonts,amssymb}
\usepackage{graphicx}
\usepackage{setspace}
\definecolor{updatecolor}{rgb}{0.5,0.0,0.8}
\newcommand{\updated}[1]{{#1}}

\title{Front-End ASIC for the STROBE-X HEMA and WFM Detectors: Concept and Design}

\author[a,b,c,*]{Gianluigi De Geronimo}
\author[d]{Paul S. Ray}
\author[d]{Eric A. Wulf}
\author[e]{Colleen A. Wilson-Hodge}
\author[f]{Eric Burns}
\author[g]{Yuri Evangelista}
\author[d]{Anthony Hutcheson}
\author[h]{Thomas J. Maccarone}
\author[i]{Gianluigi Zampa}

\affil[a]{Nuclear Engineering and Radiological Sciences, University of Michigan, Ann Arbor, MI 48109, USA}
\affil[b]{Electrical and Computer Engineering, Stony Brook University, Stony Brook, NY 11794, USA}
\affil[c]{DG Circuits, dgcircuits.com, Syosset, NY 11973, USA}
\affil[d]{Space Science Division, U.S. Naval Research Laboratory, Washington, DC 20375 USA}
\affil[e]{Astrophysics Branch, NASA's Marshall Space Flight Center, Huntsville, AL 35812, USA}
\affil[f]{Louisiana State University, Dept. of Physics and Astronomy, 202 Nicholson Hall, Baton Rouge, LA, 70803, USA}
\affil[g]{INAF - Istituto di Astrofisica e Planetologia Spaziali, Via del Fosso del Cavaliere 100, 00133, Roma, Italy}
\affil[h]{Department of Physics \& Astronomy, Texas Tech University, Box 41051, Lubbock, TX, 79409-1051, USA}
\affil[i]{Istituto Nazionale di Fisica Nucleare (INFN), Trieste TS, Italy}

\begin{document} 
\maketitle

\begin{abstract} This paper presents the NSX front-end ASIC, being developed to read charge signals from the HEMA and WFM X-ray detectors for the STROBE-X mission. The ASIC reads out signals from up to 64 anodes of linear Silicon Drift Detectors (SDDs). When unloaded, the ASIC channel has a \updated{charge resolution, expressed in Equivalent Noise Charge (ENC) of about 2.8 e$^-$}. Once connected to the SDD anode we anticipate, for the 80 keV energy range, a \updated{ENC of about 10.7 e$^-$ at a leakage current of 2 pA, which corresponds to a FWHM of about 145 eV at 6 keV once the Fano-limited statistics from charge generation in Si is included}. The acquisition is event-triggered and, for events exceeding the threshold, the ASIC measures the peak amplitude and stores it in an analog memory for subsequent readout. The ASIC can also force the measurement of the sub-threshold channels neighboring the triggered channel, including the ones that belong to neighbor chips by using bi-directional differential inter-chip communication. Alternatively, the ASIC can measure the amplitudes of all channels at the time of the first detected peak. Additional features include a high-resolution option, channel power down and skip function, a low-noise pulse generator, a temperature sensor, and the monitoring of the channel analog output and trimmed threshold. The power consumption of the individual channel is $\sim$590 µW and, when including all shared circuits, it averages to $\sim$670 µW / channel.   
\end{abstract}

% Include a list of up to six keywords after the abstract

% Include email contact information for corresponding author
{\noindent \footnotesize\text{* Gianluigi De Geronimo, E-mail: }\linkable{degeronimo@ieee.org} }

%\begin{spacing}{2}   % use double spacing for rest of manuscript

\section{Introduction}
\label{sect:intro}  % \label{} allows reference to this section

\begin{figure}[htb]
    \begin{center}
    \includegraphics[width=5.5in]{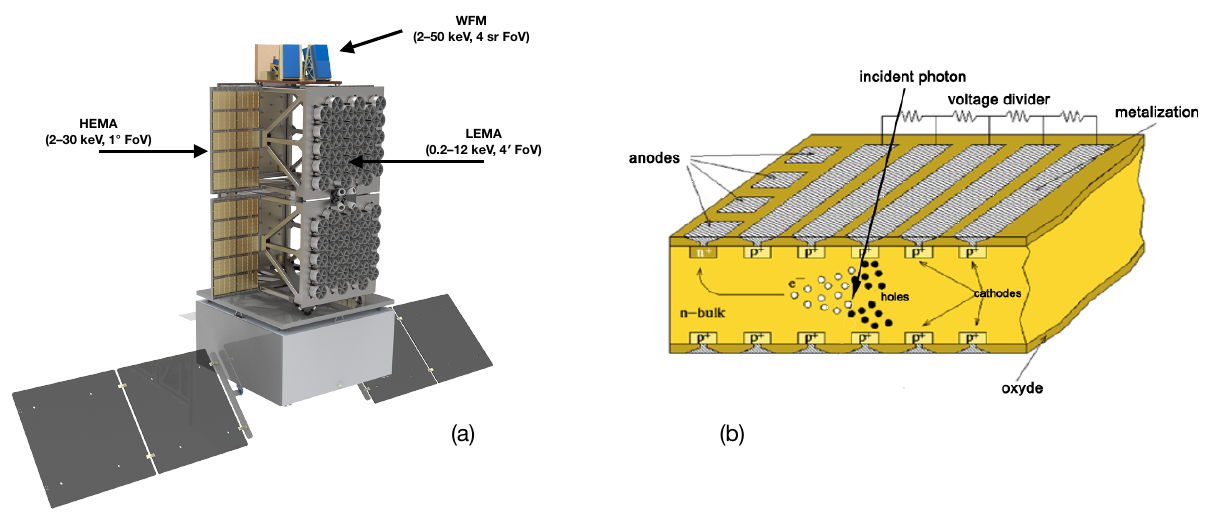}
    \end{center}
    \caption{(a) STROBE-X observatory concept showing the LEMA, HEMA, and WFM instruments and (b) working principle of the linear Silicon Drift Detector (SDD).
    \label{fig:overview}}
\end{figure}

STROBE-X (Spectroscopic Time-Resolving Observatory for Broadband Energy X-rays) is a NASA probe-class observatory designed to perform X-ray spectroscopy and timing over a broad range of energies (200 eV to 80 keV) and timescales (microseconds to years). It is part of a large multi-messenger astronomy program aiming at studying the matter in the most extreme conditions found in the universe \cite{2018SPIE10699E..19R,2019arXiv190303035R,strobex.org,JATISOverview}. As shown in Fig.~\ref{fig:overview}a, STROBE-X is composed of three instruments: a $\sim$1.6 m$^2$ soft-band (0.2 to 12 keV) Low Energy Modular Array (LEMA), a $\sim$3.4 m$^2$ hard-band (2 to 30 keV, and up to 80 keV in extended mode) High Energy Modular Array (HEMA), and a $\sim$1,456 cm$^2$ hard-band (2 to 50 keV) Wide Field Monitor (WFM). Details on the architecture and purpose of each instrument can be found in other papers in this Volume \cite{JATISOverview,JATISLEMA,JATISHEMA,JATISWFM}.

The HEMA and WFM instruments of interest to the work presented in this paper are based on linear Silicon Drift Detectors (SDDs) \cite{1984NIMPR.225..608G,1991NIMPA.306..187V,2011NIMPA.633...22C,2014JInst...9P7014R} where the anodes are arranged in a single fine pitch row with drift lines parallel to each other and the charge is collected along one edge of the sensor, as shown in Fig.~\ref{fig:overview}b. The SDD design is optimized for X-ray astronomy to achieve a 450 $\mu$m depletion with a sensitive energy range up to 80 keV.

\begin{figure}[htb]
    \begin{center}
    \includegraphics[width=5.5in]{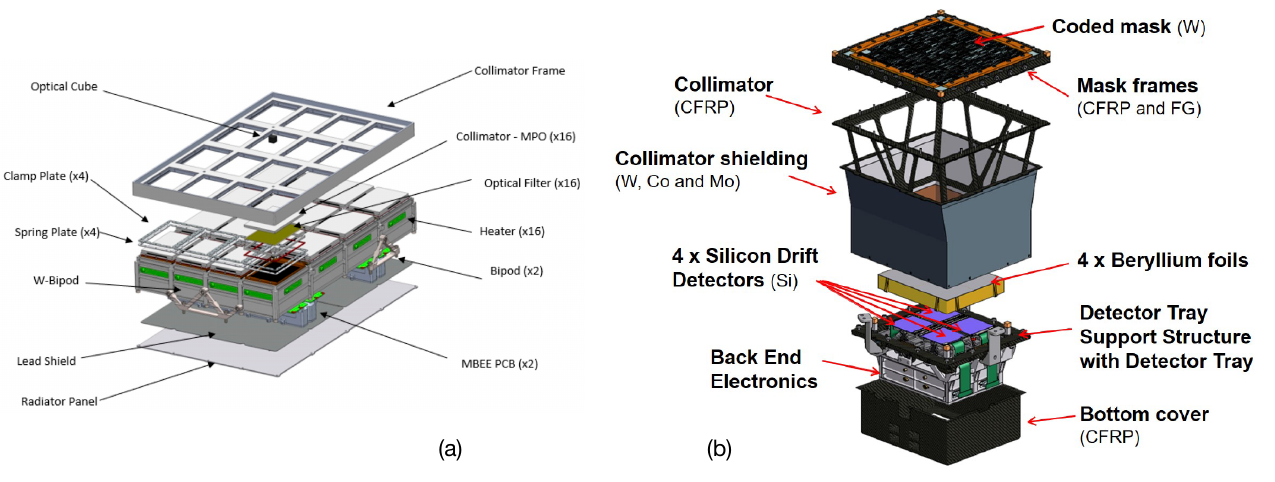}
    \end{center}
    \caption{Exploded CAD drawing of (a) an individual HEMA module and (b) and single WFM camera.
    \label{fig:modules}}
\end{figure}

Each individual HEMA SDD has an active area of 70 mm in the drift direction and 108.5 mm in the equipotential direction. The drift direction is segmented in two halves of 35 mm each with the highest voltage along the central symmetry axis, and the charge is collected at the two opposite edges with 112 anodes on each side spaced 970 $\mu$m. The estimated event multiplicity (i.e. the number of triggered adjacent strips per event from charge sharing) is between one and two. Eight 14-channel low-noise Application-Specific Integrated Circuits (ASICs) provide the readout of each half independently for a total of 16 ASICs per detector. As shown in Fig.~\ref{fig:modules}a, each individual HEMA module is composed of $4 \times 4$ of such detectors and capillary plate collimators with a 70\% open area ratio. The modules are then assembled in four panels, where each panel includes 5 $\times$ 2 = 10 modules (Fig.~\ref{fig:overview}a) for a total of 40 modules covering a total area of $\sim$4.86 m$^2$. When taking into account the collimators, the effective sensitive area is reduced to $\sim$3.4 m$^2$.

Each individual WFM SDD has a size of 70 mm in the drift direction and 64.9 mm in the equipotential direction, realizing almost a square. Like the HEMA, the drift direction is segmented in two halves of 35 mm each with the highest voltage along the central symmetry axis, and the charge is collected at the two opposite edges with 384 anodes on each side spaced 169 $\mu$m. The estimated event multiplicity is between one and eight. Six 64-channel ASICs provide the readout of each half independently for a total of 12 ASICs per detector. As shown in Fig.~\ref{fig:modules}b each individual WFM module is composed of 2 $\times$ 2 detectors, a 25 $\mu$m thick Beryllium window to prevent the direct impact of micro-meteorites and small orbital debris particles, and a coded mask. Eight WFM modules are assembled in four pairs where, for each pair, the two modules are placed orthogonal to each other in order to obtain precise two-dimensional (2D) source positions from the combined information. The total sensitive area of the 8 WFM camera modules is $\sim$1,453 cm$^2$.

With regards to the front-end ASIC for the HEMA and WFM instruments, the relevant specifications and requirements are summarized in Table~\ref{tab:specs}. The \updated{requirements on the FWHM include the Fano-limited statistics from charge generation in Si at 6 keV}. The \updated{requirements on the charge resolution, expressed in Equivalent Noise Charge (ENC), include a $\sim$20$\%$} safety margin. \updated{We anticipate that at the beginning of life the instruments will comfortably meet the requirements, while towards the end of life due to the increased leakage current from SDD degradation, low temperatures down to $-55^\circ$ combined with short peaking times may be needed, with the possible drawback of some ballistic deficit.}

\begin{table}[htb]
     \caption{Front-end electronics specifications and requirements for each instrument.}
    \label{tab:specs}   
    \centering
    \begin{tabular}{lll}
\hline
\textbf{Parameter} & \textbf{HEMA} & \textbf{WFM} \\
\hline
Anode Capacitance&	90 fF&	80 fF\\
Anode Pitch	& 970 $\mu$m &169 $\mu$m \\
Number of Anodes per Half Detector & 112 & 384 \\
Number of Electronic Channels per ASIC & 14 & 64 \\
Wirebond Interconnect Capacitance \footnotesize $^\dag$  & 40 fF & 30 fF \\
Minimum Leakage Current	& 7 pA	 & 2 pA \\
Maximum Leakage Current	& 300 pA &	90 pA \\
Triggered Anodes per Event (Event Multiplicity) & 1 to 2 & 1 to 8 \\
Maximum Event Rate per Channel \footnotesize $^\ddag$ & 15 cps & 	5 cps \\
Energy Resolution FWHM at 6 keV \updated{in Si} \footnotesize $^\diamondsuit$ &	200 eV &	300 eV \\
Equivalent Noise Charge per Channel \footnotesize $^\ddag$	& 15 e$^-$ & 13 e$^-$ \\
Maximum Power Consumption per Channel &	650 $\mu$W & 720 $\mu$W \\
Maximum Energy	      &       \multicolumn{2}{c}{up to 80 keV \footnotesize $^\S$} \\
Energy Threshold \footnotesize $^\ddag$ & \multicolumn{2}{c}{$\sim$470 eV	} \\
Charge Drift Time	  &          \multicolumn{2}{c}{$\le 2.2$ $\mu$s	}\\
Charge Collection Time at Anode	 &          \multicolumn{2}{c}{ $\le 160$ ns	}\\
Operating Temperature	&      \multicolumn{2}{c}{ $-55$ $^\circ$C to $+35$ $^\circ$C}\\
Radiation Tolerance \updated{TID}	    &        \multicolumn{2}{c}{10 krad} \cr
\hline
\footnotesize {$^\dag$ estimated values; $^\ddag$ including event multiplicity}\\
\footnotesize {$^\diamondsuit$ \updated{for single-anode} events; $^\S$ extended mode \cite{JATISHEMA}}
\end{tabular}

\end{table}

\updated{This paper is organized as follows: in Section 2 we discuss the ASIC architecture and functionality, and in Section 3 we discuss the CMOS technology, the physical layout, the planned interconnects, and we present selected post-layout transistor-level simulations.}

\section{ASIC Architecture}

\begin{figure}[htb]
     \begin{center}
     \includegraphics[width=\textwidth]{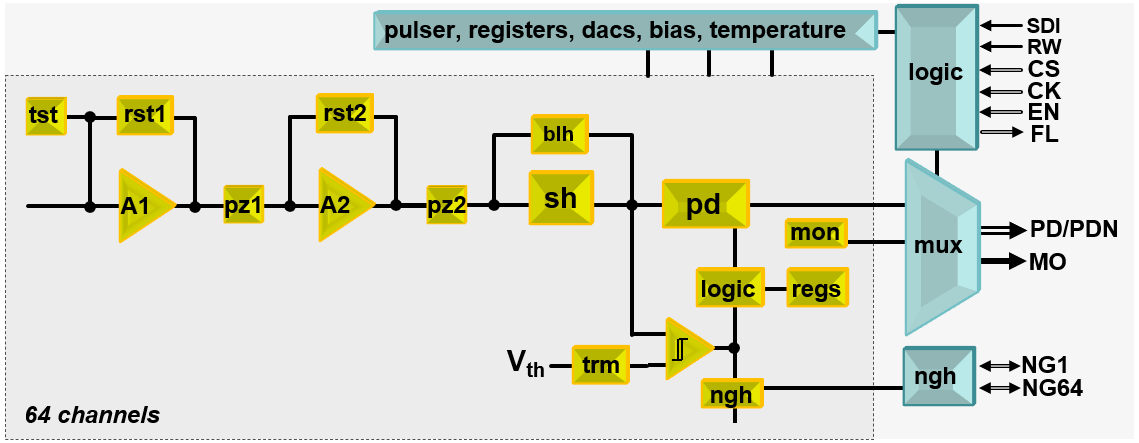}
    \end{center}
    \caption{Block diagram illustrating the NSX ASIC architecture (acronyms specified in \S 2).
    \label{fig:architecture}}
\end{figure}

The architecture of the first ASIC prototype is shown in Fig.~\ref{fig:architecture} with the individual channel enclosed in the dashed rectangle. Optimized for the WFM instrument, it includes functionality and programmability that will allow its \updated{initial} use with the HEMA instrument as well. Design modifications to optimize the ASIC for the HEMA instrument\updated{, especially to limit the power consumption,} will be done at a later time. 

The ASIC integrates 64 channels, each composed of a dual-stage low-noise charge amplifier, a shaper (sh) with baseline stabilizer, an event discriminator with trimmable threshold (trm), a peak detector (pd), control logic, and a configuration register. Shared among channels are the neighbor logic, multiplexers, bias circuits, two DACs to control the threshold and test pulse generator amplitudes, shared configuration registers, and a temperature sensor.

\subsection{Charge Amplifier}

The input MOSFET is an optimized\cite{2005ITNS...52.3223D} p-channel with gate size $L / W = 270$ nm / 40 $\mu$m (4 $\times$ 10 $\mu$m fingers) biased at a Drain current $I_\mathrm{D} = 62$ $\mu$A and a $V_\mathrm{DS} = 440$ mV (saturation voltage $V_\mathrm{DSAT} \approx 90$ mV), resulting in a gate capacitance $C_{G} \approx 73$ fF and a transconductance $g_{m} \approx 890$ $\mu$S with a sub-threshold slope coefficient $n \approx 1.33$ and an inversion coefficient $\gamma \approx 0.59$. Including the low-frequency noise, with a coefficient $K_{f} \approx 0.64 \times 10^{-24}$ J and a slope $\alpha_{f} = 1.064$ \cite{2011ITNS...58.1376D}, the MOSFET can yield a \updated{ENC of about} 2.8 e$^-$ rms without load, and a slope on the order of \updated{2.6 e$^-$} per 100 fF of input load at a peaking time of 1 $\mu$s. The user has the option of increasing the Drain current to $I_{D} = 170$ $\mu$A for a $g_{m} = 1.68$ mS yielding $\sim$2.6 e$^-$ unloaded with a slope of $\sim$3.1 e$^-$ per 100 fF of input load at a peaking time of 500 ns, which may be beneficial in case of high detector leakage current (high parallel noise), although at expenses of some ballistic deficit and the additional power consumption from the increased Drain current. For the trace, wirebond pad, and bond wire, we estimate a total capacitance on the order of 65 fF, to be added to the SDD anode capacitance. Once connected to the SDD anode, a resolution of $\sim$10.7 e$^-$ \updated{($\sim$145 eV FWHM at 6 keV once the Fano-limited statistics from charge generation in Si is included)} is anticipated \updated{at a leakage current of 2 pA} for the energy range of $\sim$80 keV (see simulations in \S 3.2), which corresponds to an analog dynamic range\updated{, defined as ratio between the maximum charge and the ENC,} in excess of 880. 

The charge amplifier (A1, rst1, pz1) implements a continuous mirror reset\cite{2008ITNS...55.1604D} with p-channel MOSFETs of size $L / W$ = 20 $\mu$m / 220 nm and $\sim$10 fF Metal-Oxide-Metal (MOM) capacitors, capable of self-adapting to leakage currents from \updated{$\sim$15 fA} to $\sim$500 pA, and providing through the pole-zero compensation (pz1)\cite{2008ITNS...55.1604D} a linear charge gain of 32. A programmable bias current of either 1 pA or 20 pA\updated{, realized using scaling mirrors,} can be enabled to emulate the leakage current from the sensor when needed (e.g. when the sensor is either not biased or not connected to the channel). The second stage of the charge amplifier (A2, rst2, pz2), similar but \updated{with the input MOSFET operating at lower bias current and with feedback} based on Metal-Insulator-Metal (MIM) capacitors, provides an additional linear charge gain programmable to either 32 or 64, resulting in \updated{a total charge gain of either 1,024 or 2,048, needed to minimize the noise contribution from the shaper and the following stages. Overall, after the shaper, a charge-to-voltage conversion gain of either $\sim$400 or $\sim$800 mV/fC is achieved} to cover, with the available voltage swing of 1.5 V, maximum energies of $\sim$84 keV and $\sim$42 keV respectively. For test purposes, an injection capacitor (tst) of 10 fF (MOM) connected to the integrated shared pulse generator can be enabled. The power consumed by the two charge amplifier stages is $\sim$110 $\mu$W.

\subsection{Shaper and Baseline Stabilizer}

The shaper (sh) is a 3\textsuperscript{rd} order with complex-conjugate poles realized using the delayed dissipative feedback (DDF) configuration\cite{2011ITNS...58.2382D}, with peaking time programmable to 0.25, 0.5, 1, and 2 $\mu$s and characterized by noise coefficients A\textsubscript{vw} = 0.85, A\textsubscript{vf} = 0.54 and A\textsubscript{iw} = 0.61 for white series, $1/f$ series and white parallel contributions respectively (unilateral noise power spectral densities)\cite{2005ITNS...52.3223D}. The baseline stabilizer is based on the baseline holder configuration\cite{2000ITNS...47..818D}, it provides a non-linear high-pass filter with a small-signal $-3$dB frequency on the order of 1 kHz, and it sets the shaper output baseline to $\sim$150 mV using a band-gap referenced voltage characterized by high rejection against process and temperature variations. Particular attention has been paid to guarantee high stability vs temperature and detector leakage current (see \updated{post-layout} simulations in \S 3.2). The power consumed by the shaper plus stabilizer is $\sim$248 $\mu$W.

\subsection{Discrimination}

The discriminator is based on a comparator, a trimmer (trm), and logic. The comparator is designed to achieve a hysteresis of about 5 mV to discriminate \updated{signals as low as} $\sim$280 eV in the 84 keV range, which is lower than the required $\sim$470 eV for worst-case event multiplicity. The trimmer allows per-channel adjustment in $\sim$800 $\mu$V steps \updated{($\sim$ 22.5 eV at the maximum conversion gain) }for a total of $\sim$12 mV (4 bit per channel). The coarse threshold can be adjusted using a shared 10-bit DAC in $\sim$800 $\mu$V steps over a range from 0 V to $\sim$820 mV. \updated{The noise contribution from the entire discriminator is on the order of 200 µV rms.} The power consumed by the comparator plus the trimmer DAC and amplifier is $\sim$38 $\mu$W.

Events that exceed the threshold are processed for peak detection as described later. The user has the option to force the acquisition and processing of the two channels neighbor to the channel exceeding the threshold. This includes channels that belong to a neighbor chip, where chips communicate event occurrence to each other using two bi-directional LVDS interfaces (one per neighbor chip)\cite{2013ITNS...60.2314D}. Additionally, the user has the option to force the acquisition and processing of all channels in the same chip once a channel exceeds the threshold. Two options allow the user to disable the discrimination either after a first channel exceeds the threshold or after a first channel finds the peak amplitude. These last two options have been introduced to limit, when needed, the impact of pick-up from digital activity.

\subsection{Peak Detection}

\begin{figure}[htb]
    \begin{center}
    \includegraphics[width=4.2in]{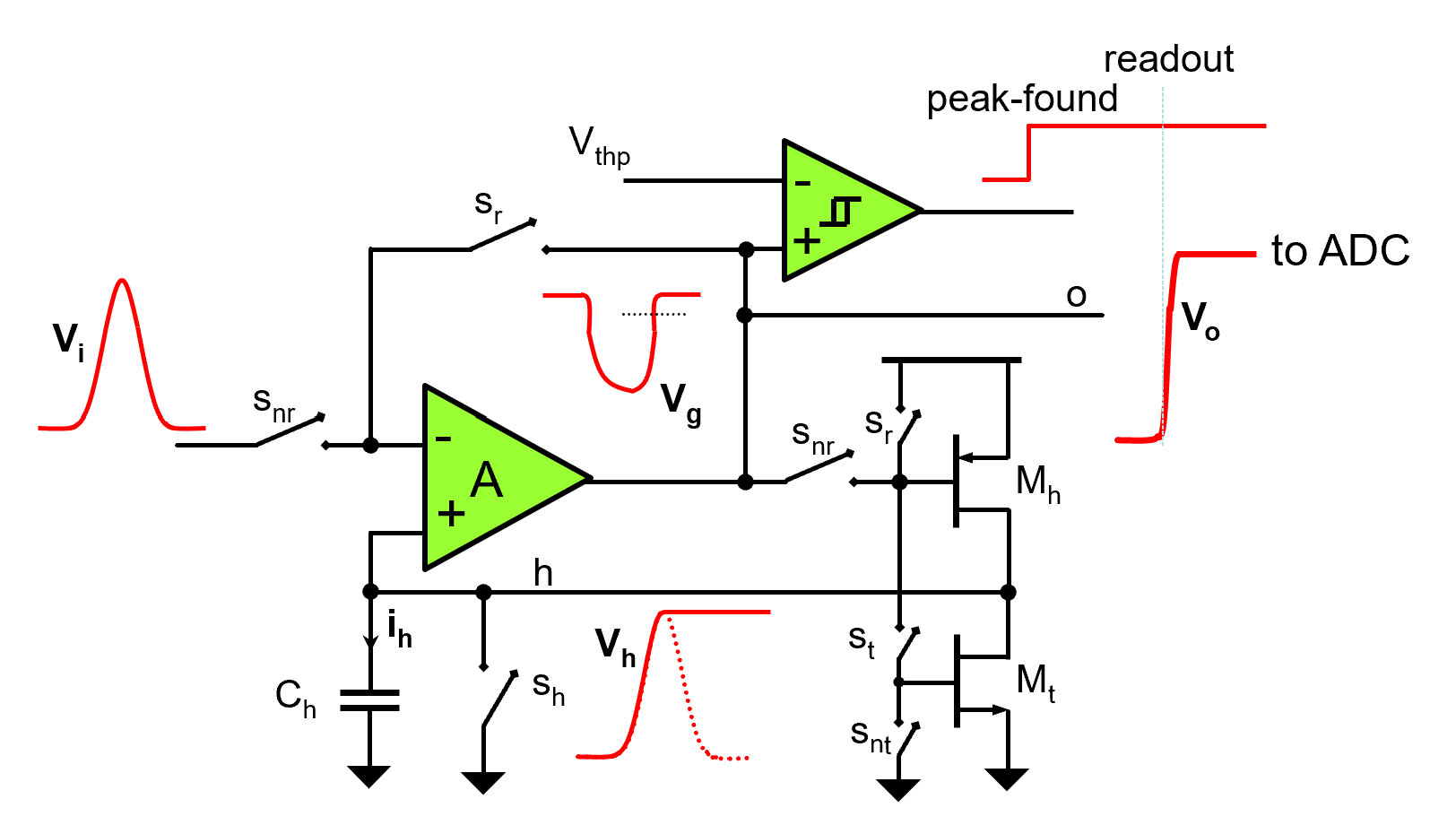}
        
    \end{center}
    \caption{Simplified schematic illustrating the three-phase operation of the peak detector circuit.
    \label{fig:peakdetect}}
\end{figure}

The peak detector (pd) provides the extraction and storage of the peak amplitude of the shaped pulse. The circuit is based on the offset-free error-cancellation concept\cite{2002NIMPA.484..544D,GDG09} and it operates in three consecutive phases, as illustrated in the simplified schematic of Fig.~\ref{fig:peakdetect}.

When the acquisition starts (i.e. after the circuit has been reset by briefly closing the switch s\textsubscript{h}), the circuit is configured for tracking-mode where the switch s\textsubscript{t} is closed (and the complement s\textsubscript{nt} open), the switches s\textsubscript{r} are open (and the complements s\textsubscript{nr} closed), and the hold node \textit{h} continuously tracks the baseline of the analog shaper, including noise. The tracking is affected by errors from the offset-voltage and finite common-mode rejection of the amplifier A. When the shaped pulse exceeds the event threshold (or, optionally, when other channels force the acquisition as discussed earlier), the circuit changes its configuration to peak-detect-mode where the switch s\textsubscript{t} is open (and the complement s\textsubscript{nt} closed) while the hold node \textit{h} continues tracking the rising part of the analog pulse until it reaches the maximum amplitude (peak). Since there is no discharge path for the node \textit{h}, the peak value is held on the hold capacitor C\textsubscript{h}. The user can program the channel to detect either the first occurring peak or the highest peak within the same acquisition time window. After the peak is found, the circuit switches its configuration into read-mode, where the switches s\textsubscript{r} are closed (and the complements s\textsubscript{nr} open) and the amplifier is reconfigured to operate as a unity-gain buffer for the readout of the voltage stored in the hold capacitor C\textsubscript{h}. While the voltage held at the node \textit{h} is affected by errors from the offset-voltage and finite common-mode rejection of the amplifier, the buffered voltage available at the output node \textit{o} has these errors cancelled. This is possible due to the re-usage of the same amplifier during all three consecutive phases \cite{2002NIMPA.484..544D}. Particular attention has been paid to guarantee high accuracy at all amplitudes, including the ones very close to the discrimination threshold (see simulations in \S 3.2).

When the peak is found, a peak-found signal is generated (see Fig. 4), OR-ed with the peak-found signals from all other channels. The result of the OR function is a shared peak-found signal occurring at the first peak-found event among all channels. Following the shared peak-found a flag is released by the ASIC with a $\sim$400 ns delay, alerting the external control logic that one or more above-threshold events have been (or are being) processed. The control logic stops the acquisition after a pre-determined time and starts the multiplexed readout, as described in the next subsection. The pre-determined time is a system-level variable controlled by the user which takes into account the maximum time difference between events of interest within the same acquisition. If enabled, an option forces all channels to store the voltage in correspondence of the first peak-found. This option has been added to allow measurement and possibly subtraction of common-mode pick-up, if present. The power consumed by the peak-detection circuit is $\sim$80 $\mu$W.

\subsection{Acquisition and Readout}

The external data acquisition system (DAQ) drives the ASIC digital control signals EN (enable), CS (chip-select) and CK (clock), and receives from the ASIC the digital signal FL (flag). The ASIC enters the Acquisition mode by asserting EN. In Acquisition mode the channels become sensitive to events. Events that exceed the threshold (and, if configured for those options, either the sub-threshold neighbors or all other sub-threshold channels) are processed for peak detection. Once a peak is found, the ASIC asserts the FL and, after a pre-determined time, the DAQ stops the Acquisition by lowering EN. Next, the ASIC enters the Readout mode by asserting CS. Once in Readout mode, the ASIC generates an internal digital token signal, which can be transferred to the first channel at the falling edge of a CK pulse. Once the token reaches the first channel, the FL becomes a channel-event indicator for that channel. If FL is high, the channel needs to be read out and the stored peak amplitude of that channel is made available at the analog output PD to be converted by an external analog-to-digital converter (ADC). A rising edge of CK will change the purpose of the FL signal, which now becomes the threshold-crossing indicator for that channel, providing information whether or not the channel exceeded the threshold. Once the ADC sampling is complete the DAQ can move the token to the next channel by providing an additional CK pulse. Channels which present a low FL are empty channels and do not need to be read out. For those empty channels, the token can stop for a much shorter time then the case where the peak amplitude needs to be read out and converted. The readout sequence is repeated until all 64 channels have been scanned and, when needed, converted. The channel identifier associated with each conversion is derived from the number of CK pulses. Once all 64 channels have been scanned, CS can be de-asserted and the ASIC enters a short acquisition-reset cycle and becomes ready for a new Acquisition-Readout sequence.

\begin{figure}[htb]
\begin{center}
        \includegraphics[width=4.0in]{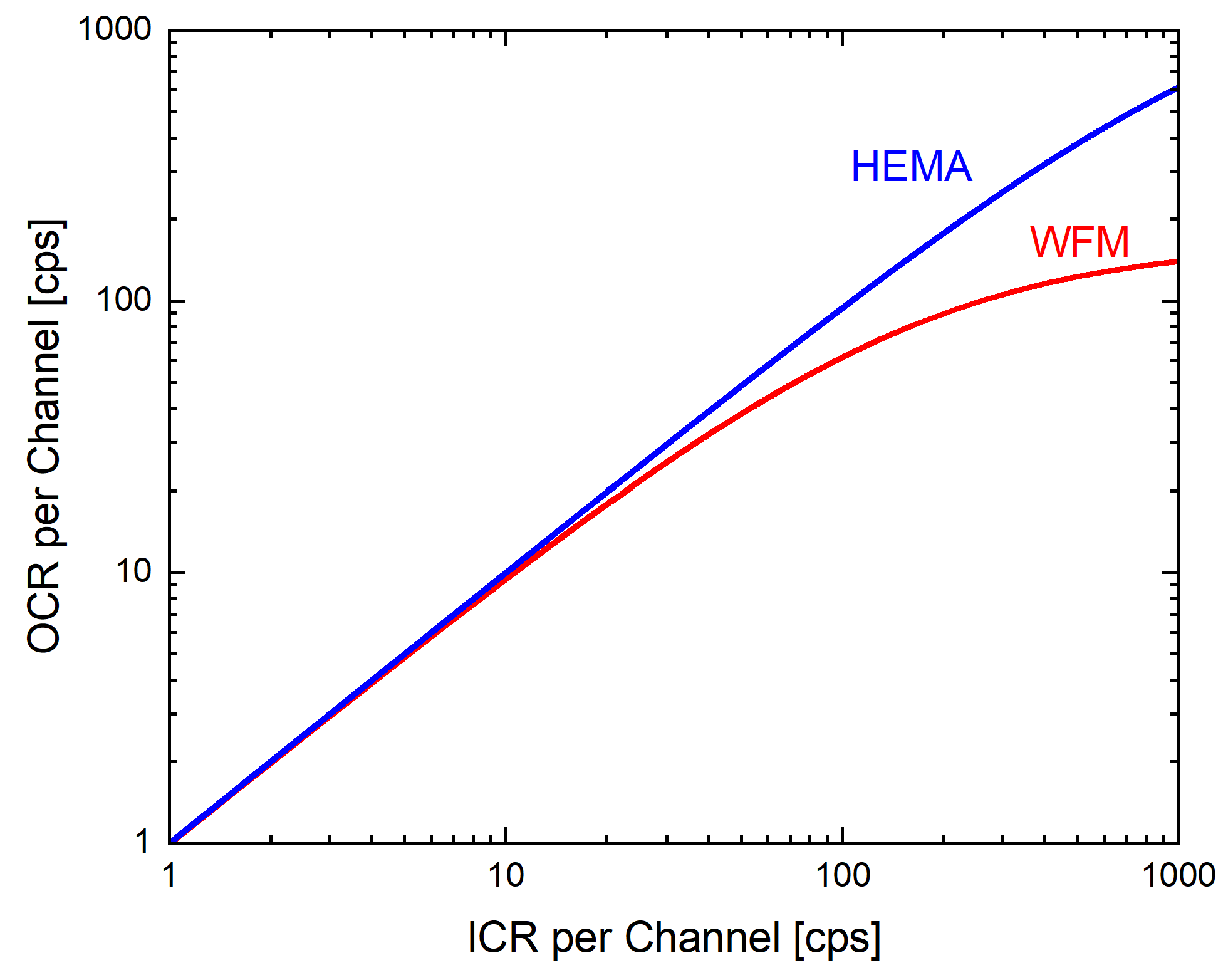}
\end{center}
    \caption{Calculated per-channel Output Count Rate (OCR) vs Input Count Rate (ICR). At the required maximum count rate, the corresponding event loss for the HEMA and WFM instrument would be 0.8$\%$ and 2.8$\%$ respectively.
    \label{fig:OCR}}
\end{figure}

Assuming conservatively an event multiplicity of 8, a 10 MHz CK, and a 1 MS/s ADC (e.g. the space-rated version of the LTC2378), a worst-case readout time on the order of 14 $\mu$s can be achieved. Considering that each half-detector operates independently and does not measure during a readout, a non-paralyzable analysis can be performed in order to calculate, given the dead time, the per-channel output count rate (OCR) vs input count rate (ICR). The result, shown in Fig.~\ref{fig:OCR}, assumes a worst-case peaking time of 2 $\mu$s, the worst-case event multiplicity, and it takes into account that the ASICs in each half detector would be read out in parallel. From Fig.~\ref{fig:OCR} it can be observed that the proposed architecture can satisfy the rate requirement for both the HEMA and WFM instruments with a maximum event loss of $\sim$0.8$\%$ and $\sim$2.8$\%$ respectively. In the extreme case where all the channels need to be read out (e.g. to subtract common-mode pick-up), the readout time will exceed 64 $\mu$s and the HEMA and WFM instruments will suffer a maximum event loss of $\sim$2.5$\%$ and $\sim$11$\%$ respectively. For such cases, ADCs with shorter conversion time may be considered.

Depending on the power budget, available CK frequency, and ADC conversion speed, architectures where ADCs are shared among multiple ASICs may be adopted. Additionally, a channel that has been powered down by configuration will be skipped by the token during the readout process, thus reducing the number of channels to be scanned. This functionality can be particularly useful when the ASIC is used to read out the signals from the HEMA detector, where only 14 channels are needed out of 64.

\subsection{Interfaces, Additional Functionality, Total Power Consumption\text-color{red}{, and Radiation Tolerance}}

The analog interface includes, along with the 64 sensitive inputs, the two analog outputs PD and MO. The peak-detector output PD allows the readout of the stored analog peak amplitudes during readout. It also includes a programmable buffer capable of driving several hundreds of pF with a sub-200 ns settling time. Optionally, the user can enable a second buffered output PDN to have a fully differential and buffered output signal. The monitor output MO can be configured to measure various internal analog signals. The user can monitor, for each individual channel, either the shaper output or the trimmed threshold. The user can also monitor the voltage of the $\sim$1.24 V integrated bandgap reference, the $\sim$150 mV analog baseline reference, the output of the 10-bit DAC which sets the untrimmed event threshold, the output of the 10-bit DAC which sets the amplitude of the integrated pulse generator, and the output of an integrated temperature sensor which has a nominal response of $\sim$924.5 mV + $\sim$3.27 mV/$^\circ$C.

The complete digital interface includes two CMOS inputs (SDI, RW) for the configuration, three LVDS inputs (CS, CK, EN) for the control of the acquisition and readout, one low-power LVDS output (FL) for the event/channel flag which requires an external 1 k$\Omega$ termination, and two low-power LVDS bi-directional lines (NG1, NG64) used by neighbor ASICs to communicate threshold events to each-other and do not require termination.

A combination of the CK, RW and CS signals allows the user to generate either a soft reset, which optionally resets all acquisition circuits in preparation of a new acquisition, or a hard reset which performs the full ASIC reset including the configuration.

In normal (non-high-resolution) mode the channel power consumption is $\sim$590 $\mu$W. In high-resolution mode the power consumption increases to $\sim$890 $\mu$W. In power down mode the channel consumption reduces to $\sim$90 $\mu$W. The shared circuits consume up to $\sim$6 mW depending on the configuration. The total power consumption in normal configuration is on the order of 43 mW (i.e. $\sim$670 $\mu$W per channel).

\updated{The configuration registers are designed using the Dual Interlocked Storage Cell (DICE) \cite{DICE}, which has high tolerance to Single Event Upsets (SEU). The anticipated Total Ionizing Dose (TID), lower than 10 krad, does not represent a concern for the selected CMOS technology \cite{TID_CMOS}.}

\section{CMOS Technology, Physical Layout and Post-Layout Simulations}

The design is in a well-established commercial mixed-signal 180 nm CMOS technology selected for its anticipated long-term availability, reliable models, and affordable fabrication costs in shared and dedicated runs. The technology has also been extensively modeled and characterized for operation at room and cryogenic temperature \cite{2011ITNS...58.1376D}. 

\subsection{Physical Layout and Interconnections}

\begin{figure}[htb]
\begin{center}
    \includegraphics[width=\textwidth]{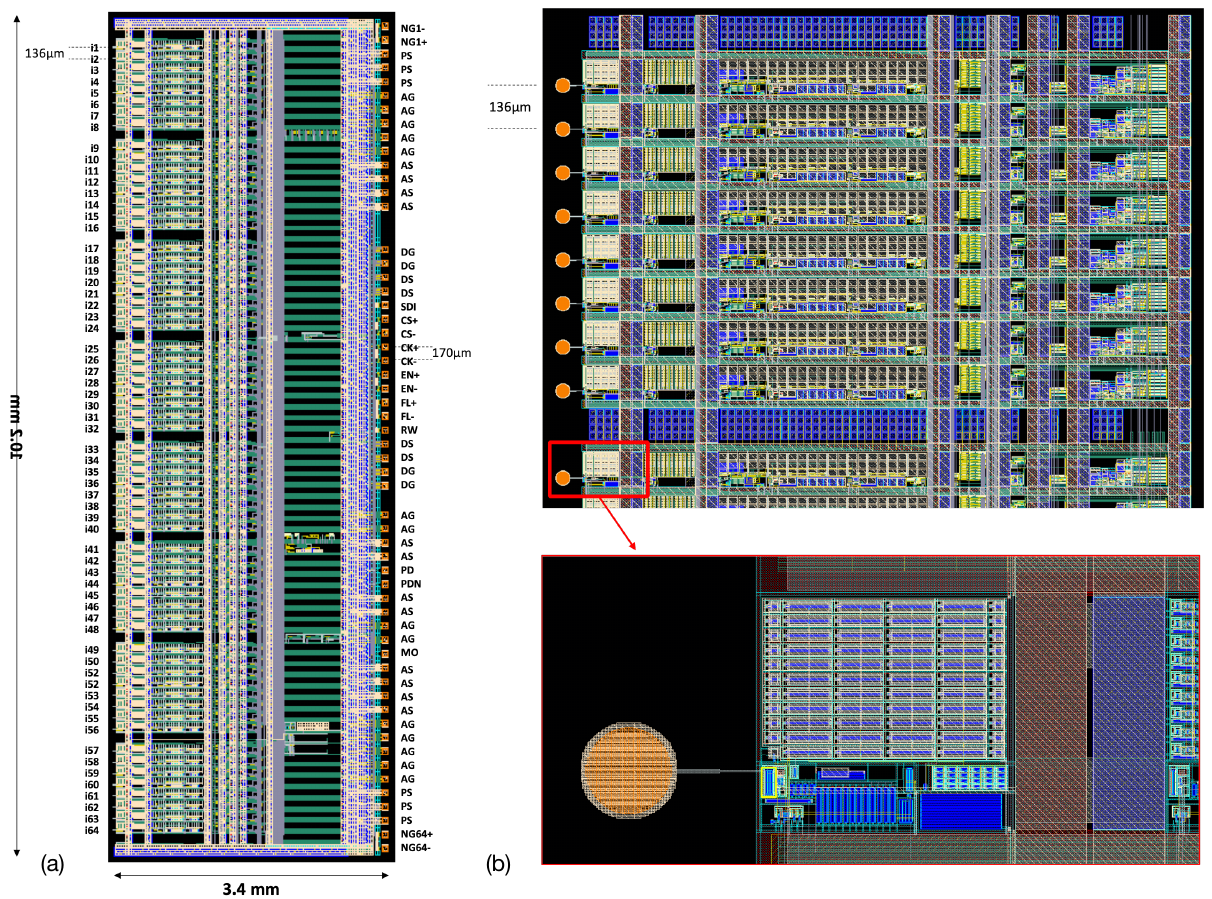}
\end{center}
    \caption{Physical layout of (a) the NSX ASIC and (b) of a group of 8 channels with zoomed detail of the input.}
    \label{fig:layouts}
\end{figure}

The physical layout of the first ASIC prototype, with a size of 3.4 mm $\times$ 10.2 mm, is shown in Fig.~\ref{fig:layouts}a. The aspect ratio is within the 3:1 recommended to guarantee die integrity during die saw. The 64 inputs are located on the left side of the die, while the rest of the interface (supplies, grounds, analog and digital signals) is located on the right side.

Each channel measures 1.9 mm $\times$ 136 $\mu$m plus the interconnect and round-shaped input pad (Fig.~\ref{fig:layouts}b). The parasitic capacitance of the pad, characterized by an opening diameter of 40 $\mu$m suitable for 1 mil bond wire and realized using the two uppermost metal layers (M5 and M6), plus the metal interconnect is on the order of 25 fF. The channels are grouped in eight (Fig.~\ref{fig:layouts}b) with a pad pitch of 136 $\mu$m and inter-group spacing rails of 136 $\mu$m. The spacing rails and the two top and bottom rails are used to bring the charge amplifier supply from the right side to the left side of the ASIC. In Fig.~\ref{fig:layouts}b the input MOSFET is also visible, surrounded by a yellow rectangle.

The right side of the ASIC layout is dedicated to the shared circuits and to pads for the supplies, grounds, analog and digital interfaces, with a minimum pad pitch of 170 $\mu$m. The uppermost and lowermost differential pads (NG) provide the bi-directional inter-chip communication for neighboring as described in the previous section. The PS pads provide the dedicated 1.8 V for the charge amplifier supply (Source of the input MOSFET), the AS and AG pads provide the 1.8 V to 0 V analog supply and ground. The digital interface is localized at the center, surrounded by the DS and DG pads which provide the 1.8 V to 0 V digital supply and ground.

\begin{figure}[htb]
\begin{center}
        \includegraphics[width=5.5in]{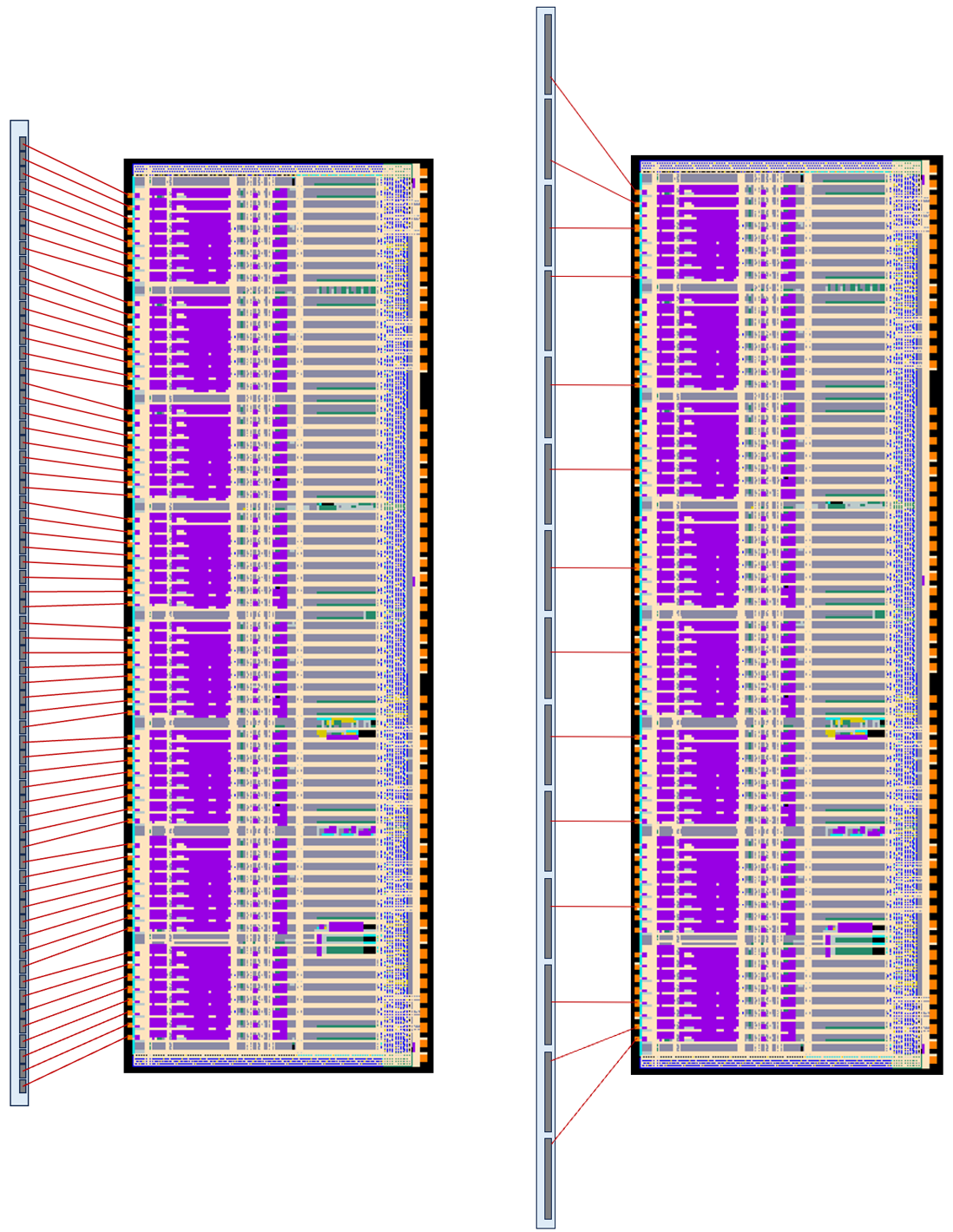}
\end{center}
    \caption{Planned wire bonding drawing for the WFM case (left, 64 anodes at 169 $\mu$m pitch) and\updated{, for initial testings, the} HEMA case (right, 14 anodes at 970 $\mu$m pitch). A maximum wire bond angle of 45$^\circ$ is maintained.}
    \label{fig:bonding}
\end{figure}

Fig.~\ref{fig:bonding} shows the planned wirebond drawing for both the WFM \updated{case and, for initial testings, the HEMA case}. In the WFM case, where the anode pitch is 169 $\mu$m, the spacing between adjacent chips covering 64 SDD anodes will be in excess of 500 $\mu$m, after subtracting $\sim$30 $\mu$m for the seal ring and $\sim$80 $\mu$m for the scribe line (saw street). In the HEMA case, where the anode pitch is 970 $\mu$m, the spacing between adjacent chips covering 14 SDD anodes will be on the order of 3.2 mm. The die will be positioned in order to have a maximum bond wire angle of 45 degrees and a maximum bond wire length of $\sim$3.6 mm, with an estimated worst case parasitic capacitance on the order of 40 fF.

\subsection{Transistor-Level Simulations}

\begin{figure}[htb]
\begin{center}
        \includegraphics[width=4.0in]{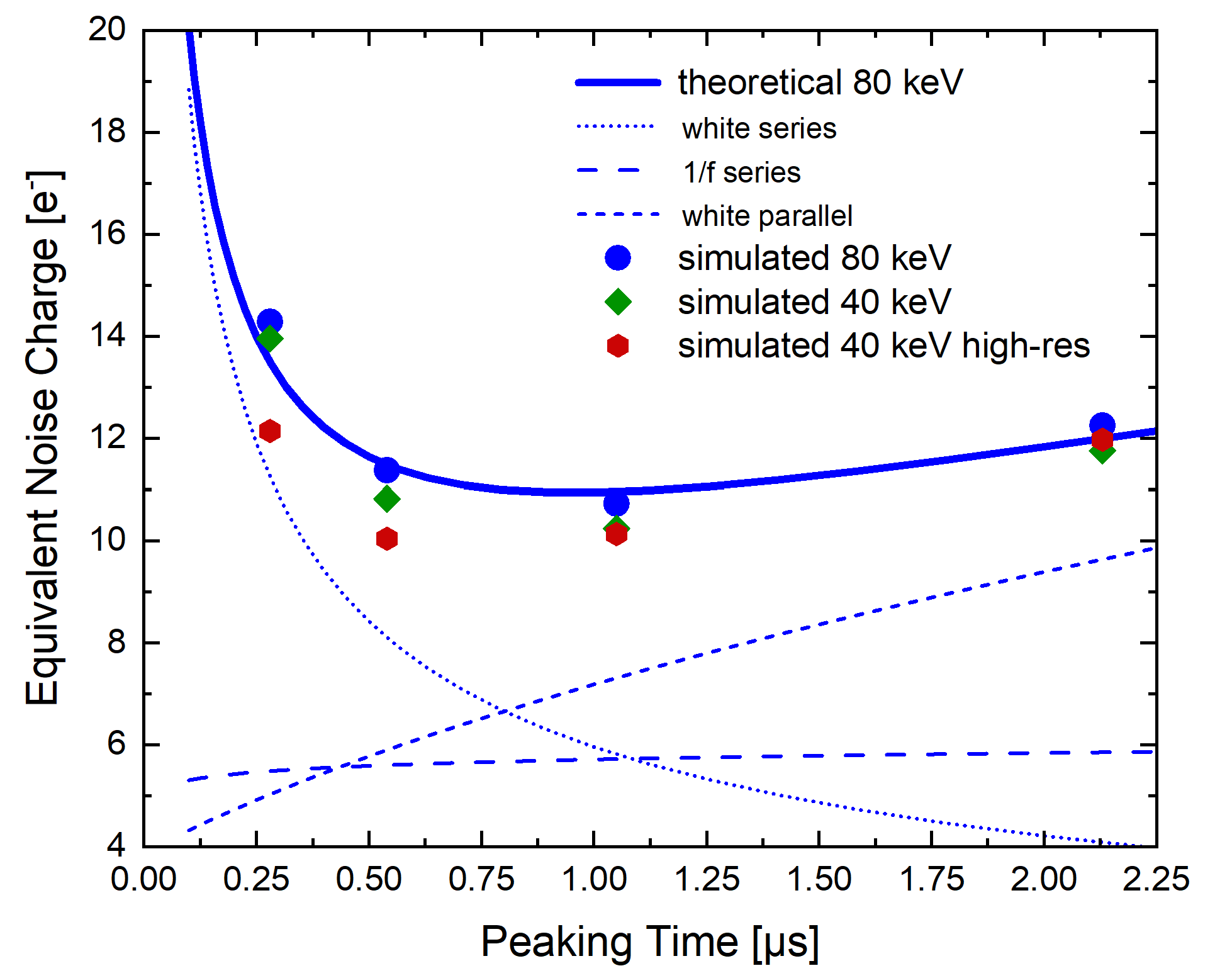}
\end{center}
    \caption{Theoretical and simulated Equivalent Noise Charge for the WFM case at the nominal leakage current.}
    \label{fig:enc}
\end{figure}

The design has been extensively simulated transistor-level using the industry-standard CAD tools from Cadence Design Systems, with pre- and post-layout netlists where the latter includes the extracted parasitic capacitances associated with the physical layout. Fig.~\ref{fig:enc} shows the extracted ENC at the four peaking times for the energy ranges of the 80 keV, 40 keV and 40 keV high-res (i.e. with the high-resolution mode enabled). This simulation assumes the WFM case and a \updated{leakage current} of 2 pA. The shot noise associated with the detector leakage current is included. For comparison, the theoretical ENC is also shown, along with the three dominant contributions (white series, $1/f$ series, and white parallel). The theoretical curves include the noise contributions from the stages following the charge amplifier, and which amounts to $\sim$5\% of the total ENC.

\begin{figure}[htb]
\begin{center}
    \includegraphics[width=\textwidth]{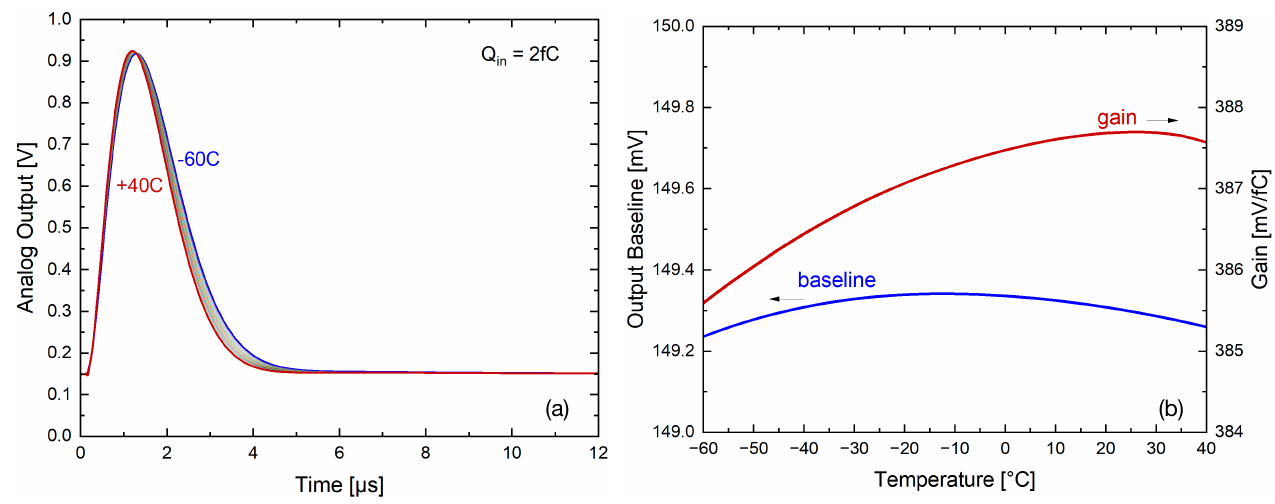}
\end{center}
    \caption{Simulated analog response to 2 fC (a) and baseline and conversion gain (b) vs temperature.}
    \label{fig:resp-vs-temp}
\end{figure}

The simulated analog response vs temperature to an input charge $Q\textsubscript{in} = 2$ fC (signal current modeled using an approximate Dirac  $\delta$ function) is shown in Fig.~\ref{fig:resp-vs-temp}a, along with the conversion gain and output baseline (Fig.~\ref{fig:resp-vs-temp}b), for temperatures from $-60$ $^\circ$C to $+40$ $^\circ$C. A gain change on the order of 0.5\% can be observed over the entire temperature range.

\begin{figure}[htb]
\begin{center}
    \includegraphics[width=6.0in]{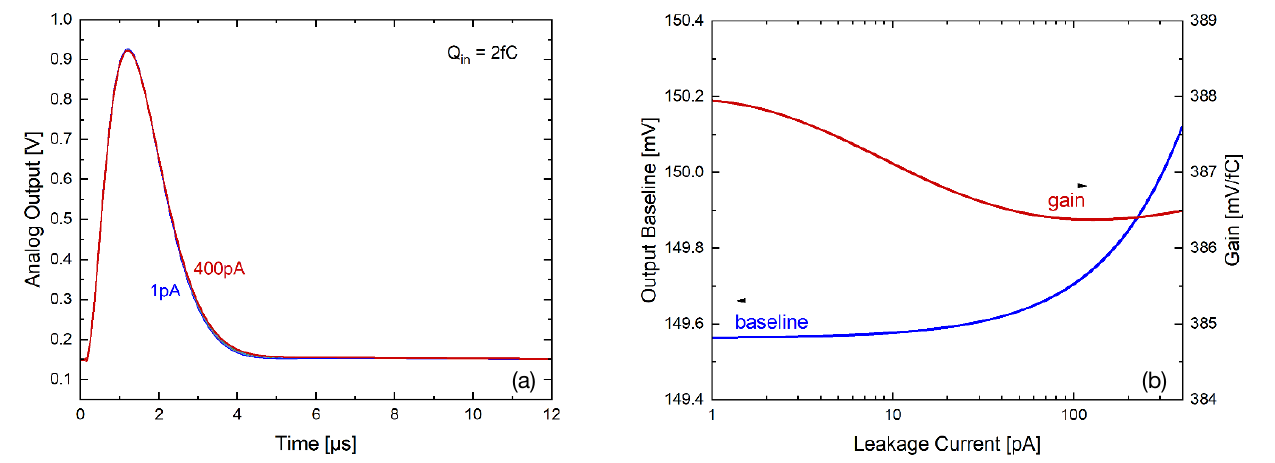}
\end{center}
    \caption{Simulated analog response to 2fC (a) and baseline and gain (b) vs detector leakage current.}
    \label{fig:resp-vs-leakage}
\end{figure}

Similarly, the simulated analog response vs detector leakage current to an input charge $Q\textsubscript{in} = 2$ fC ($\sim$delta Dirac current) is shown in Fig.~\ref{fig:resp-vs-leakage}a, along with the gain and output baseline (Fig.~\ref{fig:resp-vs-leakage}b), for leakage currents from 1 pA to 400 pA.

\begin{figure}[htb]
    \begin{center}
        \includegraphics[width=6.0in]{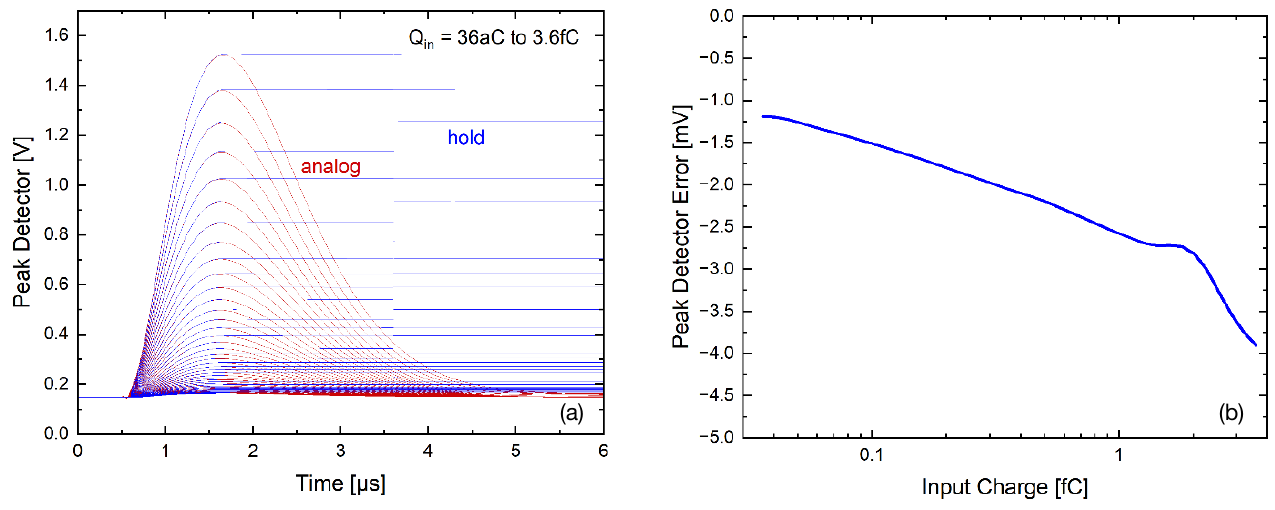}
    \end{center}
    \caption{Simulated peak detector response (a) and accuracy (b) vs input charge Q\textsubscript{in}.}
    \label{fig:resp-vs-charge}
\end{figure}

In Fig.~\ref{fig:resp-vs-charge}a the simulated peak detector response and accuracy are shown at 1 $\mu$s peaking time, for input charges from 36 aC ($\sim$810 eV) to 3.6 fC ($\sim$81 keV). A relative error on the order of 2.75 mV can be observed, which corresponds to less than 0.2\% over the full range. Taking into account the linear component, an error well below 0.1\% can be expected.

\begin{figure}[htb]
\begin{center}
        \includegraphics[width=5.0in]{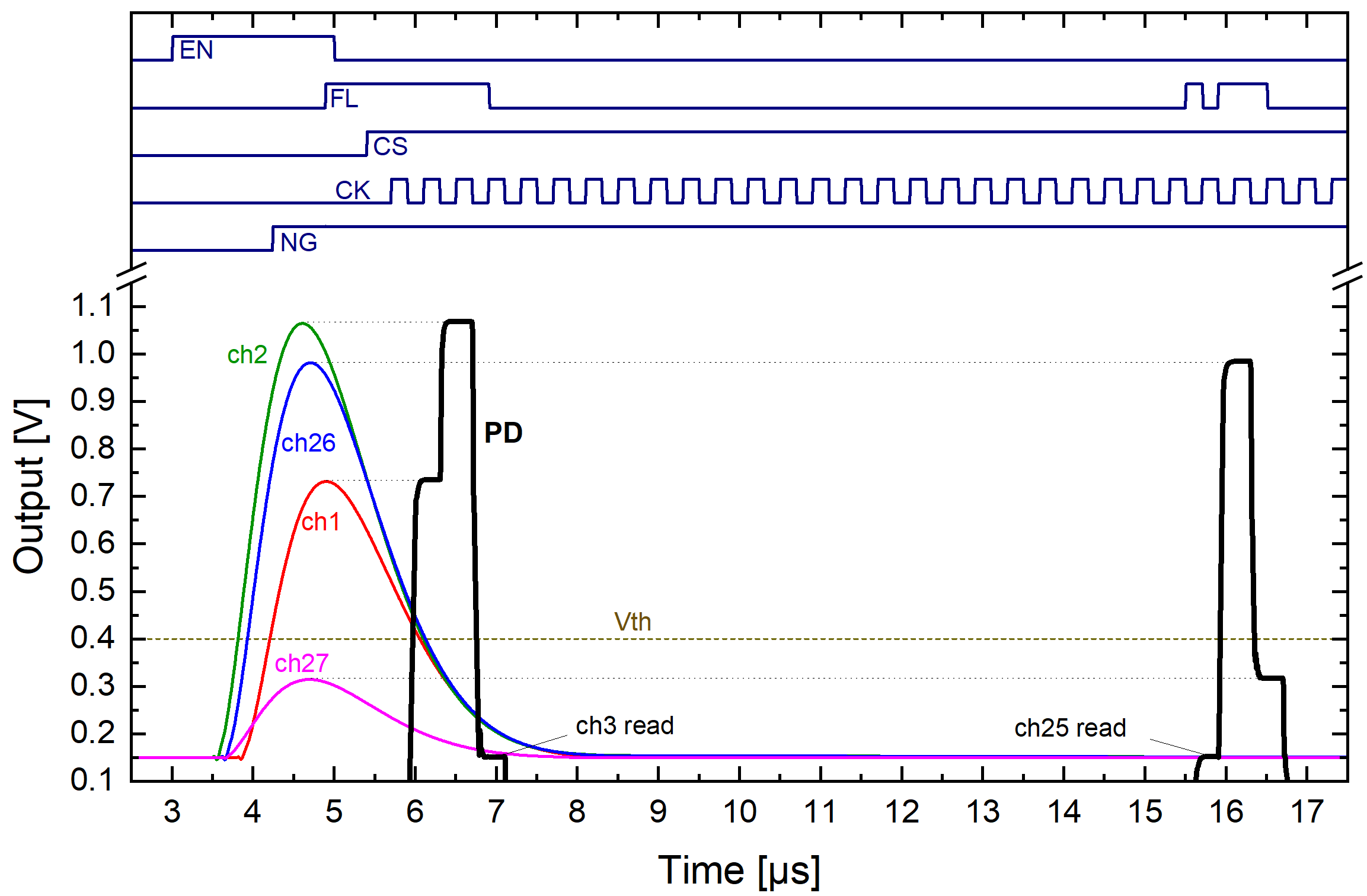}

\end{center}
    \caption{Example of a full acquisition and readout cycle.}
    \label{fig:readout}
\end{figure}

A simulation example of a complete acquisition and readout sequence is shown in Fig.~\ref{fig:readout}. In this example the discrimination threshold is set to $\sim$400 mV, and events occur in channels 1, 2, 10 and 11 where the latter does not exceed the threshold. The signals EN, FL, CS, CK and NG (inter-chip) are also shown, in digital form. For this simplified example, a single readout clock frequency of 2.5 MHz has been adopted.

The sequence starts with setting EN high. Once the events occur and the first peak is found the ASIC releases, with a $\sim$400 ns delay, the FL. The DAQ de-asserts EN, asserts CS and generates the sequence of CK. At the first falling edge of CK the high FL indicates that channel 1 needs to be converted and the PD output presents its peak amplitude. At the next rising edge of CK the high FL indicates that channel 1 exceeded the threshold. At the second CK cycle the same happens for channel 2. At the third CK cycle the sub-threshold neighbor channel 3 is read out, with amplitude close to the baseline, and at the falling edge of CK the FL is de-asserted indicating that channel 3 did not exceed the threshold. The next 21 CK cycles indicate no events until channels 25, 26, and 27 where both 25 and 27 did not exceed the threshold. The inter-chip neighboring signal occurs at the threshold crossing of channel 1.

\appendix    % this command starts appendixes

% \disclosures 
\subsection*{Disclosures}
The authors have no potential conflicts of interest to disclose.

\subsection* {Code, Data, and Materials Availability} 
Data sharing is not applicable to this article, as no new data were created or analyzed.

\subsection*{Acknowledgments}

The authors are very grateful to Marco Feroci (National Institute of Astrophysics (INAF), Italy), Valter Bonvicini (National Institute for Nuclear Physics (INFN), Italy), and Jean in ’t Zand (Netherlands Institute for Space Research (SRON), The Netherlands) for valuable discussions and help with specific requirements. 
The authors are also very grateful to Keith Gendreau (NASA Goddard Space Flight Center, Greenbelt, MD, USA), Adam Goldstein (Universities Space Research Association, Huntsville, AL, USA), Peter Jenke (University of Alabama in Huntsville, Huntsville, AL, USA), and Henric Krawczynski (Washington University in St. Louis, St. Louis, MO, USA).

This work was supported by NASA proposal number 22-APRA22-0026.

%%%%% References %%%%%

\bibliography{report}   % bibliography data in report.bib
\bibliographystyle{spiejour}   % makes bibtex use spiejour.bst

%%%%% Biographies of authors %%%%%

\vspace{2ex}\noindent\textbf{Gianluigi De Geronimo} is an adjunct professor with the University of Michigan and with the Stony Brook University, and a consultant. He received his MS and PhD degrees in Electronics and Communications from Milan Polytechnic. He is the author of more than 120 journal papers and one book chapter. His current research interests include low-noise front-end ASIC design and characterization. He is member of IEEE and editor for IEEE TNS.

\vspace{2ex}\noindent\textbf{Paul S. Ray} is an astrophysicist at the U.S. Naval Research Laboratory. He received his A.B. in physics from the University of California, Berkeley in 1989 and his Ph.D. in physics from Caltech in 1995. His current research interests include multiwavelength studies of pulsars and other neutron star systems and X-ray instrument and mission development. He is a member of AAS, IAU, and Sigma Xi.

\vspace{2ex}\noindent\textbf{Eric A. Wulf} is a research physicist with the Naval Research Laboratory. He received his MS degree in Physics from the University of Virginia and his PhD degree in Physics from the Duke University.  He is the author of more than 110 journal papers. His current research interests include data acquisition and nuclear physics detectors. He is member of IEEE and of the American Astronomical Society.

\vspace{2ex}\noindent\textbf{Colleen Wilson-Hodge} is an astrophysicist at NASA's Marshall Space Flight Center. She received her MS degree in  1996 and PhD degree in 1999. Both degrees were in Physics from the University of Alabama in Huntsville. Her current research interests include multimessenger transients along with X- and gamma-ray instrument and mission development. She is a member of AAS and IAU.

\vspace{1ex}
\noindent Biographies and photographs of the other authors are not available.

%%% FIGURES

%\end{spacing}
\end{document}